\documentclass[12pt]{article}
\usepackage{amsmath,amssymb,amsfonts,amscd,bbm,youngtab,graphicx}
\begin{document}

%%%%%%%%%%%%%%%%%%%%%%%%%%%%%%%%%%%%%%%%%%%%%%%%%%%%%%%%%%%%%%%%%%%%%
%                     Titel Page                                    %
%%%%%%%%%%%%%%%%%%%%%%%%%%%%%%%%%%%%%%%%%%%%%%%%%%%%%%%%%%%%%%%%%%%%%

\thispagestyle{empty}
\renewcommand{\thefootnote}{\fnsymbol{footnote}}
\setcounter{footnote}{1}

\vspace*{-1.cm}
\begin{flushright}
OSU-HEP-04-3
\end{flushright}
\vspace*{1.8cm}

\centerline{\Large\bf Chiral Gauge Models for Light}
\vspace*{3mm}
\centerline{\Large\bf Sterile Neutrinos}

\vspace*{18mm}

\centerline{\large\bf 
K.S. Babu\footnote{E-mail: \texttt{babu@okstate.edu}} and 
Gerhart Seidl\footnote{E-mail: \texttt{gseidl@susygut.phy.okstate.edu}}}
      
\vspace*{5mm}
\begin{center}
{\em Department of Physics, Oklahoma State University,}\\
{\em Stillwater, OK 74078, USA}
\end{center}

\vspace*{20mm}

\centerline{\bf Abstract}
\vspace*{2mm}
We construct a family of simple gauge models in which
three sterile neutrinos become naturally light by virtue of a generalized seesaw
mechanism involving a chiral gauge symmetry. Examples where the chiral gauge
group is $SU(5)'$, $SU(7)'$, $SU(3)'$ and/or their descendants are presented.
A unified model based on $SO(10)\times SO(10)'$ which embeds many of these models is constructed wherein three light sterile neutrinos are just as natural
as the three ordinary neutrinos. These gauge models have relevance to current neutrino oscillation data, including the LSND anomaly.

%\keywords{Keyword1; keyword2; keyword3.}

\renewcommand{\thefootnote}{\arabic{footnote}}
\setcounter{footnote}{0}

\newpage

%%%%%%%%%%%%%%%%%%%%%%%%%%%%%%%%%%%%%%%%%%%%%%%%%%%%%%%%%%%%%%%%%%%%%
%                     Introduction                                  %
%%%%%%%%%%%%%%%%%%%%%%%%%%%%%%%%%%%%%%%%%%%%%%%%%%%%%%%%%%%%%%%%%%%%%
\section{Introduction}
Over the past few years, solar \cite{solar}, atmospheric \cite{atmospheric},
reactor \cite{KamLAND}, and accelerator \cite{K2K} neutrino
experiments have remarkably improved our knowledge of the neutrino mass and mixing parameters. Specifically, solar and atmospheric neutrino data are now
excellently understood within a three-neutrino oscillation scheme, where the
neutrino mass squared splittings are respectively
$\Delta m_\odot^2\simeq 7.5\times 10^{-5}\:{\rm eV}^2$ and
$\Delta m_{\rm atm}^2\simeq 2.0\times 10^{-3}{\rm eV}^2$ \cite{barger}.
However, the evidence for $\overline{\nu}_\mu-\overline{\nu}_e$ oscillations
found by the Liquid Scintillator Neutrino Detector (LSND) experiment at
Los Alamos \cite{LSND}, which  will soon be tested by 
the ongoing MiniBooNE experiment at Fermilab \cite{MiniBooNE}, would demand
a third mass squared difference
$\Delta m^2_{\rm LSND}\gtrsim 10^{-1}\:{\rm eV}^2$, which cannot be
accommodated in a three-neutrino oscillation scenario. The LSND anomaly indicates instead
the presence of $n\geq 1$ additional neutrinos with masses of order
$\sim 1\:{\rm eV}$, which give rise to a (3+n) neutrino oscillation scheme
providing additional mass squared splittings. This new species of neutrinos cannot couple to the $Z$ boson, and hence must be sterile with respect to weak interactions. Although a (3+1) neutrino mass scheme \cite{valle,bar} seems now
to be almost ruled out
\cite{giun}, a combined fit
of the short-baseline experiments Bugey \cite{Bugey}, CCFR \cite{CCFR84},
CDHS \cite{CDHS}, CHOOZ \cite{CHOOZ}, KARMEN \cite{KARMEN}, and LSND
shows, that the LSND signal can become compatible with the other neutrino
oscillation data sets with two (or more) light sterile neutrinos in a (3+2) neutrino mass scheme \cite{sorel}.

In models for (3+2) neutrino oscillations
\cite{Babu:2003is,McDonald:2004pa}, it is important to
provide a rationale for the smallness of sterile neutrino masses, since the
usual seesaw mechanism \cite{yana79} does not explain why a sterile neutrino $\nu'$ would be light. Actually, if the effective low-energy theory is the Standard Model
(SM), then there is no reason why $\nu'$ would not acquire a mass of the order of some high cutoff scale $\Lambda\simeq 10^{13}-10^{19}\:{\rm GeV}$.
Indeed, there exists a number of suggestions to realize light sterile
neutrinos \cite{babu,Foot:1995pa,Babu:2003is,McDonald:2004pa}. By copying the criteria which make the seesaw
mechanism successful to the sterile sector, we have constructed in
Ref.~\cite{Babu:2003is} the simplest anomaly-free chiral gauge theory
which naturally leads to the (3+2) neutrino oscillation scheme. The sterile sector of this
model consists of an $SU(2)'$ gauge group with the $\nu'$ transforming as
a spin 3/2 multiplet where symmetry breaking is achieved by a single spin 3/2 Higgs field. In this paper, we wish to generalize
this idea to a larger class of chiral gauge symmetries in order to arrive at
models for $(3+n)$ neutrino oscillations in which $n$ sterile neutrinos are
light.

Our basic approach is here to extend the gauge group of the Standard Model (SM)
$G_{SM}=SU(3)_c\times SU(2)_L\times U(1)_Y$ by some
``sterile'' gauge group $G'$ so that the full gauge symmetry is
$G_{SM}\times G'$, with the $\nu'$ transforming as a chiral
representation\footnote{By this we mean a fermionic representation for which
mass terms are forbidden by gauge invariance.} of $G'$. All SM particles have zero $G'$ charges. In all
our constructions, we require $G'$ to be a gauge symmetry, rather than a global symmetry, since presumably only gauge symmetries will survive quantum
gravity corrections. In this way, the $\nu'$ are protected from acquiring large explicit masses of order
$M_{Pl}\simeq 10^{19}\:{\rm GeV}$ and therefore can serve as candidates for naturally light sterile neutrinos. For $3+n$ light neutrinos, we will suppose
$3+n$ heavy neutrinos which are total gauge singlets of $G_{SM}\times G'$.
This would then lead to a generalization of the seesaw mechanism to the sterile sector. In analogy with the electroweak symmetry breaking in the SM,
we assume that $G'$ is spontaneously broken around the TeV scale by a suitable Higgs field
$S$, which has no direct coupling of the type $\nu'\nu'S$. To keep the situation simple, we take $S$ to be a singlet under $G_{SM}$ and require a minimal Higgs sector: a single Higgs $S$ breaks $G'$ to one of its subgroups
 and provides simultaneously sterile neutrino masses, analogous to the SM Higgs doublet.

Notice that in the special case when $G'$ becomes a copy of $G_{SM}$, we arrive at the well-studied scenario for ``mirror'' neutrinos \cite{Foot:1995pa}.
Here, however, we are interested in a chain of models where $G'=SU(N)'$, starting with fermionic fields in the following class:
\begin{equation}
 \Yboxdim{12pt}\Yvcentermath1\yng(1,1)\oplus
 (N-4)\times\overline{\Yvcentermath1\yng(1)}\:.\label{eq:Class(a)}
\end{equation}
This describes an $SU(N)'$ gauge theory with one fermion multiplet in
the antisymmetric second rank tensor representation and $N-4$ fermions in the
antifundamental representation. For $N\geq 5$, this
class of gauge theories is known to be chiral and anomaly-free and has
been analyzed extensively in the context of dynamical supersymmetry breaking
\cite{dine95,intril94}. We will also study, albeit in less detail, the chain of anomaly-free chiral gauge theories arising from fermionic fields transforming under $SU(N)'$ as 
\begin{equation}\label{eq:Class(b)}
\Yboxdim{12pt} \Yvcentermath1\yng(2)\oplus(N+4)\times
\overline{\Yvcentermath1\yng(1)}\:.
\end{equation}
Here, one fermion multiplet transforms under the
symmetric second rank tensor representation of $SU(N)'$ and $N+4$ fermionic fields
transform under the antifundamental representation.
 For $N\leq 4$, this also gives anomaly-free
gauge theories which may, however, be vectorlike. This is indeed the case
for $N=2$, for which the $SU(2)'$ spin 1 representation
{\tiny \Yvcentermath1\yng(2)} is vectorial. The simplest such theory is if $N=3$, which we analyze in some detail.

Starting from the types of gauge theories given in Eq.~(\ref{eq:Class(a)})
(and Eq.~(\ref{eq:Class(b)})),
we can easily construct for $N\geq 5$ (and $N=3$) various
new anomaly-free chiral models by simply decomposing the chiral multiplets
into irreducible representations of the subgroups of $SU(N)'$. Thus we are able to generate a whole family of chiral gauge models for light sterile neutrinos. We will also see that many of these models can be embedded into a unified theory based on $SO(10)\times SO(10)'$.

The rest of the paper is organized as follows. In Sec.~\ref{sec:SU(5)'}, we construct gauge models based on the sterile gauge symmetry $SU(5)'$ as well as its various descendants. In Sec.~\ref{sec:SO(10)}, we present a fully unified model
based on $SO(10)\times SO(10)'$. In Sec.~\ref{sec:othermodels}, we describe other models arising from $SU(7)'$ as well as $SU(3)'$ with a symmetric tensor representation. Finally, in Sec.~\ref{sec:summary}, we give a summary of our main results and list ways of
testing these models.

\section{$G_{SM}\times SU(5)'$ model and its descendants}\label{sec:SU(5)'}
\begin{figure}
\begin{center}
\includegraphics*[bb = 192 616 412 746]{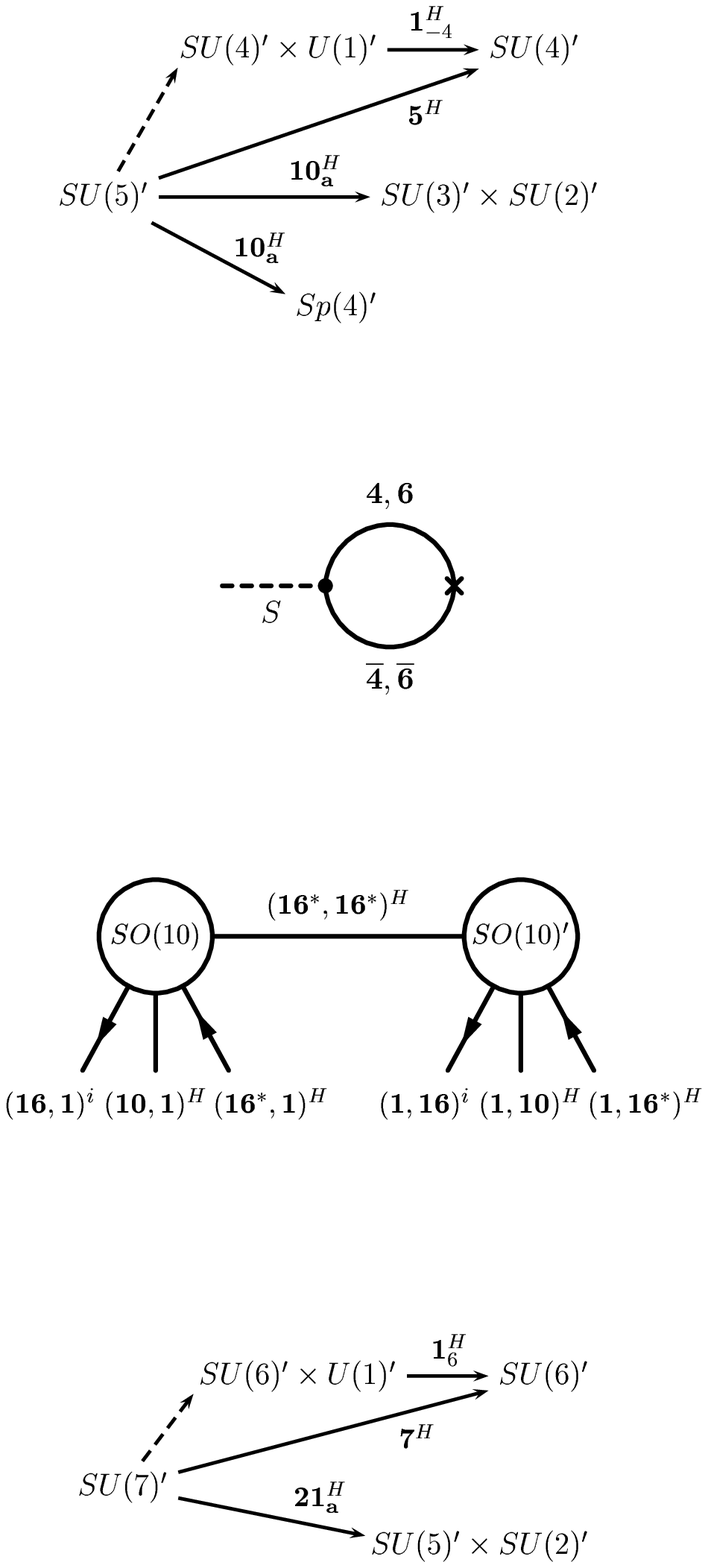}
\vspace*{-2mm}
\caption{\small Symmetry breaking in $SU(5)'$. The solid arrows represent the
symmetry breakings by suitable Higgs representations considered in the text. The dashed arrow
indicates the embedding of $SU(4)'\times U(1)'$ into $SU(5)'$ as a
maximal subgroup.}\label{fig:SU(5)'}
\end{center}
\end{figure}

In this section, we present the simplest chiral gauge models obtained from the chain shown in Eq.~(\ref{eq:Class(a)}). These models are based on an $SU(5)'$ ``sterile'' gauge symmetry or one of its descendants. The different patterns of
symmetry breaking considered in this section are summarized
in Fig.~\ref{fig:SU(5)'}.
\subsection{$SU(5)'$ model}\label{sec:SU(5)'5H}
The simplest anomaly-free chiral $SU(N)'$ gauge theory which admits fermion
representations in the chain of Eq.~(\ref{eq:Class(a)}) is $SU(5)'$.
As a simple extension of the
SM to an anomaly-free chiral gauge theory, we will therefore consider the
gauge group $G_{SM}\times SU(5)'$ with $n$ copies
(or ``generations'') of SM singlet fermions which
transform under $G_{SM}\times SU(5)'$ as
\begin{equation}\label{eq:SU(5)content}
({\bf 1},{\bf 10})^i
+({\bf 1},\overline{\bf 5})^i,
\end{equation}
where $i=1,\ldots ,n$ is the generation index of the SM singlet fermions. In
analogy with the SM, we will, in what follows, choose for definiteness $n=3$.
Notice
in Eq.~(\ref{eq:SU(5)content}), that the sterile sector then actually
becomes a copy of the usual $SU(5)$ model. This model can thus be realized as a descendant of a unified $SU(5)\times SU(5)'$ model.
While
$({\bf 1},{\bf 10})^i$ and $({\bf 1},\overline{\bf 5})^i$ in
Eq.~(\ref{eq:SU(5)content}) are sterile with respect to $G_{SM}$, the SM
particles are singlets under $SU(5)'$. We denote the SM Higgs by $H$ and the
SM lepton doublets by $\ell_i$, where $i=1,2,3$ is the generation index.
To generate small neutrino masses via the seesaw mechanism, we assume six
right-handed neutrinos $\nu^c_k$, where $k=1,\ldots, 6$, which are total
singlets under $G_{SM}\times SU(5)'$. We suppose that $SU(5)'$ is
spontaneously broken
at the TeV scale by a single SM singlet Higgs representation $S$ which
transforms as $S\sim({\bf 1},{\bf 5})^H$ under $G_{SM}\times SU(5)'$. When $S$
acquires a vacuum expectation value (VEV), $\langle S\rangle\simeq\mathcal{O}({\rm TeV})$,
$SU(5)'$ is broken down to $SU(4)'$, thereby eating 9 Nambu Goldstone bosons from $S$ via the Higgs mechanism. This is, {\it e.g.}, immediately seen
in the unitary gauge, where $\langle S\rangle$ can always be written as
$\langle S\rangle=(0,0,0,0,|s|)^T$. Since we impose minimality of the Higgs sector, $SU(4)'$ will remain unbroken.

The most general renormalizable Lagrangian relevant for neutrino masses in this model is given by
\begin{eqnarray}\label{eq:SU(5)Lagrangian}
 \mathcal{L}_Y&=&a_{i k}H\ell_i\nu^c_k+b_{ik}S
({\bf 1},\overline{\bf 5})^i\nu^c_k
+c_{ij}S^\ast({\bf 1},{\bf 10})^i({\bf 1},\overline{\bf 5})^j\nonumber\\
&+&d_{ij}S({\bf 1},{\bf 10})^i({\bf 1},{\bf 10})^j
+M_{kl}\nu^c_k\nu^c_l+{\rm h.c.},
\end{eqnarray}
where $i,j=1,2,3$ and $k,l=1,\ldots,6$. In Eq.~(\ref{eq:SU(5)Lagrangian}), the
coefficients $a_{ik}$, $b_{ik}$, $c_{ij}$, and $d_{ij}$ denote complex order one Yukawa couplings and $M_{kl}\simeq\mathcal{O}(\Lambda)$,
where $\Lambda\simeq 10^{13}-10^{19}\:{\rm GeV}$ is some high cutoff scale of the
theory. When $S$ acquires a VEV along its fifth component, the Yukawa interactions
in Eq.~(\ref{eq:SU(5)Lagrangian}) with coefficients $c_{ij}$ and $d_{ij}$
generate Dirac masses $\sim\mathcal{O}({\rm TeV})$ for
14 of the 15 fermions in each generation of $SU(5)'$, which thus
decouple from the low-energy theory below
$\langle S\rangle\simeq\mathcal{O}({\rm TeV})$. This is analogous to the
embedding of the SM in $SU(5)$ where all particles except $\nu_i$ from
$(\overline{\bf 5},{\bf 1})^i$ acquire masses from analogous Yukawa couplings. The remaining Weyl fermions, one per family (denoted as $\nu'_i$), on the
other hand, mix with the right-handed neutrinos through the interaction
$\sim b_{ik}S({\bf 1},\overline{\bf 5})^i\nu^c_k$. The $\nu^c_k$ fields have Majorana masses of order $\Lambda$, and therefore generate in the effective theory for the $\nu'$ fields small masses from non-renormalizable operators obtained after integrating out the heavy states $\nu_k^c$. The effective Lagrangian
for neutrino masses becomes
\begin{equation}\label{eq:effective}
\mathcal{L}_{\rm eff}=
\frac{Y^a_{ij}}{\Lambda}{H}^2\ell_i\ell_j
+\frac{Y^b_{ij}}{\Lambda}H S\ell_i({\bf 1},\overline{\bf 5})^j
+\frac{Y^c_{ij}}{\Lambda} S^2 ({\bf 1},\overline{{\bf 5}})^i
({\bf 1},\overline{{\bf 5}})^j+{\rm h.c.},
\end{equation}
where $Y^a_{ij}$, $Y^b_{ij}$, and $Y^c_{ij}$ are complex order one
Yukawa couplings which are related to the parameters
$a_{ij}$, $b_{ik}$, and $M_{kl}$ in Eq.~(\ref{eq:SU(5)Lagrangian}). Inserting
the VEV $\langle S\rangle$ into Eq.~(\ref{eq:effective}),
we thus observe that $\mathcal{L}_{\rm eff}$ gives rise to a seesaw mass
operator which generates per generation one active and one sterile neutrino
mass in the $\sim 10^{-2}\:{\rm eV}$ range. Since the
active and sterile neutrinos exhibit a nonzero mixing through the second term
in Eq.~(\ref{eq:effective}), we hence obtain a (3+3) scheme for sterile
neutrino oscillations.

\subsection{``Flipped'' $SU(5)'$ model}\label{sec:SU(5)'10H}
Let us now consider a variation of the model in Sec.~\ref{sec:SU(5)'5H} where only the Higgs content in the sterile sector is modified. As gauge group we
therefore have again $G_{SM}\times SU(5)'$ with 3 extra generations of SM
singlet fermions transforming according to the $G_{SM}\times SU(5)'$
representations as given
in Eq.~(\ref{eq:SU(5)content}). Like in the model in Sec.~\ref{sec:SU(5)'5H},
we require six right-handed neutrinos $\nu^c_k$, where $k=1,\ldots ,6$, which are total singlets of $G_{SM}\times SU(5)'$. We now assume that the scalar
sector is augmented by a single Higgs field $S$ which transforms as
$S\sim ({\bf 1},{\bf 10})^H$ under $G_{SM}\times SU(5)'$ and acquires a VEV 
$\langle S\rangle\simeq \mathcal{O}({\rm TeV})$.
The renormalizable neutrino mass and mixing terms of this model then read
\begin{equation}\label{eq:10HLagrangian}
  \mathcal{L}_Y=a_{i k}H\ell_i\nu^c_k+b_{ik}S^\ast
({\bf 1},{\bf 10})^i\nu^c_k
+c_{ij}S({\bf 1},\overline{\bf 5})^i({\bf 1},\overline{\bf 5})^j
+M_{kl}\nu^c_k\nu^c_l+{\rm h.c.}.
\end{equation}
where $i,j=1,2,3$ and $k,l=1,\ldots,6$. In Eq.~(\ref{eq:10HLagrangian}), the
coefficients $a_{ik}$, $b_{ik}$, and $c_{ij}$ denote complex order one
Yukawa couplings and $M_{kl}$ the cutoff scale.
For a range of parameters, $S$ will acquire a VEV of the
skew-symmetric form
$\langle S\rangle\sim \rm{diag}(0,0,0,1\otimes{\rm i}\sigma^2)$,
thereby breaking $SU(5)'$ down to
$SU(3)'\times SU(2)'$ \cite{li74}. This model resembles the flipped
$SU(5)$ model for the SM sector \cite{Barr:1981qv},
and hence we use the terminology ``flipped'' $SU(5)'$ model. Under
$SU(5)'\supset SU(3)'\times SU(2)'$, the representations
 in Eq.~(\ref{eq:SU(5)content}) decompose as
\begin{equation}\label{eq:SU(3)'xSU(2)'content}
({\bf 1},{\bf 10})^i=
({\bf 1},{\bf 3}, {\bf 2})^i
+({\bf 1},\overline{{\bf 3}},{\bf 1})+({\bf 1},{\bf 1},{\bf 1})^i,\quad
({\bf 1},\overline{\bf 5})^i=({\bf 1},\overline{\bf 3},{\bf 1})^i+
({\bf 1}, {\bf 1},{\bf 2})^i.
\end{equation}
If we define, like in the usual $SU(5)$ model, the components
$({\bf 1},{\bf 1},{\bf 2})^i\equiv(e_i',\nu_i')^T$, we observe
that the Yukawa interaction
$c_{ij}S({\bf 1},\overline{\bf 5})^i({\bf 1},\overline{\bf 5})^j$
in Eq.~(\ref{eq:10HLagrangian}) will finally generate a mass term of the form
$\sim f_{ij}(\nu_i'e_j'-\nu_j'e'_j)$, where
$f_{ij}=-f_{ji}$ due to Fermi statistics. Since the matrix $f_{ij}$ has rank
two, this interaction
will give masses $\sim\mathcal{O}({\rm TeV})$ to four out of six
states in the multiplets $(e'_i,\nu'_i)^T$ which consequently decouple from
the theory. Only one linear combination
of $e'_i$ and one linear combination of $\nu_i'$ remain massless. Absence of
$SU(2)'$ Witten anomaly \cite{witt82} also requires that one $SU(2)'$ doublet
must remain massless. Moreover, all the states from $({\bf 1},{\bf 10})^i$ and the
$({\bf 1},\overline{\bf 3},{\bf 1})^i$ state from $({\bf 1},\overline{\bf 5})^i$ will also remain massless.

These particles will however acquire masses from the dynamics of the unbroken $SU(3)'$ and $SU(2)'$. The one-loop beta function coefficients \cite{pol73}
for $SU(3)'$ and $SU(2)'$ are respectively
$-7{g_3'}^3/(16\pi^2)$ and
$-{g'_2}^3/(4\pi^2)$. In computing these coefficients, we included contributions from all the light fermionic states, and assumed that none of the scalar Higgs components from $S$ have masses below a TeV. We see that both
$SU(3)'$ and $SU(2)'$ are asymptotically
free. All massless non-trivial $SU(3)'\times SU(2)'$ representations can
hence decouple from the low-energy theory by acquiring masses through chiral symmetry breaking condensates. Note, in addition, that these states have zero
mixing with the active neutrinos.

After integrating out the heavy states $\nu^c_k$, the effective Lagrangian
relevant for neutrino masses becomes therefore similar to the Lagrangian in
Eq.~(\ref{eq:effective}) with $({\bf 1},{\bf 5})^H$ replaced by
$({\bf 1}, \overline{\bf 10})^H$ and $({\bf 1},\overline{\bf 5})^i$ replaced by
$({\bf 1},{\bf 10})^i$.
At low energies, we hence identify the 
$G_{SM}\times SU(3)'\times SU(2)'$ singlets
$({\bf 1},{\bf 1}, {\bf 1})^i$ in Eq.~(\ref{eq:SU(3)'xSU(2)'content}) as
three light sterile neutrinos giving in total a (3+3) model of
neutrino oscillations.

\subsection{$Sp(4)'$ model}\label{sec:Sp(4)'}
Let us now suppose the same gauge group and particle content as in the
``flipped'' $SU(5)'$ model in Sec.~\ref{sec:SU(5)'10H}, but consider a different symmetry
breaking of $SU(5)'$. In particular, we assume now a range of
parameters in the scalar potential, for which $S\sim({\bf 1},{\bf 10})^H$ acquires a VEV
$\langle S\rangle\simeq\mathcal{O}({\rm TeV})$ and is of the skew-symmetric
form
$\langle S\rangle\sim{\rm diag}(1\otimes {\rm i}\sigma^2,1\otimes {\rm i}\sigma^2,0)$. This VEV breaks
$SU(5)'\rightarrow Sp(4)'\sim SO(5)$ \cite{li74} such that the representations
in Eq.~(\ref{eq:SU(5)content}) decompose under $SU(5)'\supset Sp(4)'$ as
\begin{equation}\label{eq:Sp(4)'content}
({\bf 1},{\bf 10})^i=
({\bf 1},{\bf 5})^i+
({\bf 1},{\bf 4})^i
+({\bf 1},{\bf 1})^i,\quad
({\bf 1},\overline{\bf 5})^i=({\bf 1},{\bf 4})^i
+({\bf 1},{\bf 1})^i.
\end{equation}
Here, the Higgs field $({\bf 1}, {\bf 10})^H$ decomposes under $SU(5)'\supset Sp(4)'$
according to the first equation in Eq.~(\ref{eq:Sp(4)'content}) with the index ``$i$'' replaced by ``$H$''. The Yukawa Lagrangian relevant for neutrino masses is given by Eq.~(\ref{eq:10HLagrangian}). Inserting the representations in
Eq.~(\ref{eq:Sp(4)'content}) into Eq.~(\ref{eq:10HLagrangian}), we thus obtain for this model in the language of the unbroken $G_{SM}\times Sp(4)'$ subgroup the renormalizable Lagrangian for neutrino masses
\begin{eqnarray}\label{eq:Sp(4)'Lagrangian}
  \mathcal{L}_Y&=&a_{i k}H\ell_i\nu^c_k+b_{ik}({\bf 1},{\bf 1})^H
({\bf 1},{\bf 1})^i\nu^c_k
+c_{ij}({\bf 1},{\bf 1})^H({\bf 1},{\bf 4})^i({\bf 1},{\bf 4})^j\nonumber\\
&+&c_{ij}({\bf 1},{\bf 1})^H({\bf 1},{\bf 1})^i({\bf 1},{\bf 1})^j
+M_{kl}\nu^c_k\nu^c_l+{\rm h.c.},
\end{eqnarray}
where the states $({\bf 1},{\bf 1})^i$ and $({\bf 1},{\bf 4})^i$ 
belong to the decomposition of the $SU(5)'$ representations
$({\bf 1},{\bf 10})^i$ and
$({\bf 1},\overline{\bf 5})^i$ in Eq.~(\ref{eq:Sp(4)'content}), respectively. Defining $({\bf 1},{\bf 4})^i$ appearing in
Eq.~(\ref{eq:Sp(4)'Lagrangian}) in component form as
$({\bf 1},{\bf 4})^i\equiv(\psi_1^i,\psi_2^i,\psi_3^i,\psi_4^i)^T$,
we see that the third term in Eq.~(\ref{eq:Sp(4)'Lagrangian}) will generate a
mass term of the form $\sim c_{ij}(\psi_1^i\psi_2^j-\psi_1^j\psi_2^i+
\psi_3^i\psi_4^j-\psi_3^j\psi_4^i)$, where $c_{ij}=-c_{ji}$ due to Fermi statistics. Since the matrix
$c_{ij}$ has rank two, this term will give masses to eight out of twelve
components in the representations $({\bf 1},{\bf 4})^i$, which then
decouple. Moreover, we can assume that after spontaneous symmetry breaking
(SSB) the $G_{SM}\times Sp(4)'$ representations $({\bf 1},{\bf 5})^H$ and $({\bf 1},{\bf 4})^H$ acquire masses $\sim\mathcal{O}({\rm TeV})$ and thus also decouple from the theory. As a consequence,
the one-loop beta
function coefficient of $Sp(4)'$ is given by\footnote{The index $\frac{1}{2}$ of the
$Sp(4)'$ spinor representation
${\bf 4}$ can be found from the regular embedding
$Sp'(4)\supset SU(2)'\times SU(2)'$ with branching rule
${\bf 4}=({\bf 2},{\bf 1})+({\bf 1},{\bf 2})$. Note also, that a $\bf 5$
decomposes as ${\bf 5}=({\bf 2},{\bf 2})+({\bf 1},{\bf 1})$.} 
$-23{g'}^3/(48\pi^2)$. Therefore, $Sp(4)'$ is asymptotically free and
the three fundamental and four spinor representations of $Sp(4)'$,
which do not acquire masses from SSB,
decouple through confinement. Hence, after integrating out the heavy states
$\nu^c_k$, the effective Lagrangian for neutrino masses becomes similar to the Lagrangian in
Eq.~(\ref{eq:effective}) with $({\bf 1},{\bf 5})^H$ replaced by
$({\bf 1},{\bf 1})^H$ and $({\bf 1},\overline{\bf 5})^i$ replaced
by $({\bf 1},{\bf 1})^i$ (from $({\bf 10},{\bf 1})^i$ in
Eq.~(\ref{eq:Sp(4)'content})) of $G_{SM}\times Sp(4)'$. We are thus left with
one light sterile neutrino per generation, which mixes with the active neutrinos, thereby leading in total to a (3+3) neutrino oscillation scheme. Note that
one linear
combination of $({\bf 1},{\bf 1})^i$ from $({\bf 1},\overline{\bf 5})^i$ will also remain light since $c_{ij}$ has rank two, but this state has zero mixing with the other light neutrinos.

\subsection{$SU(4)'\times U(1)'$ model}\label{sec:SU(4)'}
Under the subgroup $SU(4)'\times U(1)'$ of
$SU(5)'$ with $U(1)$ generator $T=(1,1,1,1,-4)$ the representations in
Eq.~(\ref{eq:SU(5)content}) decompose as
\begin{equation}\label{eq:SU(4)content}
({\bf 1},\mathbf{10})^i
=({\bf 1},\mathbf{6}_2)^i+({\bf 1},\mathbf{4}_{-3})^i,\quad
({\bf 1},\overline{\mathbf{5}})^i=
({\bf 1},\overline{\mathbf{4}}_{-1})^i+
({\bf 1},\mathbf{1}_4)^i,
\end{equation}
where in the parenthesis $({\bf 1},{\bf x}_y)$, the subscript $y$ denotes the 
$U(1)'$ charge of the states in $({\bf 1},{\bf x}_y)$ and $i=1,2,3$. Let us
now assume for
the gauge symmetry of our model $G_{SM}\times SU(4)'\times U(1)'$ with three
generations of sterile fermions transforming according to
Eq.~(\ref{eq:SU(4)content}). Note that this gauge theory is automatically
anomaly-free since it is obtained from the model in
Sec.~(\ref{sec:SU(5)'5H}) by restriction to a subgroup, while the
$U(1)'$ charge ensures that the model is chiral. Like in
Sec.~\ref{sec:SU(5)'5H}, we furthermore assume six right-handed neutrinos
$\nu_k^c$, where $k=1,\ldots, 6$, which are total singlets of
$G_{SM}\times SU(4)'\times U(1)'$. To generate small sterile neutrino masses,
we add to the SM scalar sector a single Higgs field $S$ which transforms
under $G_{SM}\times SU(4)'\times U(1)'$ as $S\sim({\bf 1}, {\bf 1}_{-4})^H$
and breaks $SU(4)'\times U(1)'$ down to $SU(4)'$ by acquiring a VEV
$\langle S\rangle\simeq \mathcal{O}({\rm TeV})$. The most general
renormalizable Lagrangian for neutrino masses then reads
\begin{eqnarray}\label{eq:SU(4)Lagrangian}
 \mathcal{L}_Y&=&a_{i k}H\ell_i\nu^c_k+b_{ik}S
({\bf 1},{\bf 1}^i_{4})\nu^c_k
+c_{ij}S({\bf 1},{\bf 6}_2)^i({\bf 1},{\bf 6}_2)^j\nonumber\\
&+&d_{ij}S^\ast({\bf 1},{\bf 4}_{-3})^i({\bf 1},\overline{\bf 4}_{-1})^j
+M_{kl}\nu^c_k\nu^c_l+
{\rm h.c.},
\end{eqnarray}
where $i,j=1,2,3$ and $k,l=1,\ldots,6$. In Eq.~(\ref{eq:SU(5)Lagrangian}), the
coefficients $a_{ik}$, $b_{ik}$, $c_{ij}$, and $d_{ij}$ denote complex
$\mathcal{O}(1)$ Yukawa couplings and $M_{kl}\simeq\mathcal{O}(\Lambda)$.
When $S$ assumes its VEV, all non-trivial representations of $SU(4)'$ acquire
masses of order TeV at tree level and hence decouple from
the low-energy theory.\footnote{Note that the singlet in the $SU(4)'$ tensor
product ${\bf 6}\times{\bf 6}$ is in the symmetric representation.}
After integrating out the heavy states $\nu^c_k$, we
are thus left
with an effective neutrino mass Lagrangian which is similar to the Lagrangian given in Eq.~(\ref{eq:effective}) with $({\bf 1},{\bf 5})^H$ replaced by
$({\bf 1},{\bf 1}_{-4})^H$ and $({\bf 1},\overline{\bf 5})^i$ replaced by
$({\bf 1},{\bf 1}_4)^i$.
We therefore obtain one light sterile neutrino per generation,
leading to a (3+3) model of sterile neutrino oscillations.

\begin{figure}
\begin{center}
 \includegraphics*[bb=254 474 359 563]{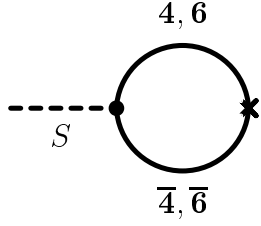}
\vspace*{-2mm}
\caption{\small Tadpole diagram generating a nonzero VEV for the $SU(4)'$
singlet $S\sim ({\bf 1},{\bf 1}_{-4})^H$ in presence of the
condensates
$\langle({\bf 1},{\bf 4}_{-3})^i({\bf 1},\overline{\bf 4}_{-1})^j\rangle$
and/or $\langle({\bf 1},{\bf 6}_{2})^i({\bf 1},\overline{\bf 6}_{-2})^j\rangle$
.}
\label{fig:tadpole}
\end{center}
\end{figure}

It is instructive to examine the effect of the condensates on the
$U(1)'$ symmetry breaking. Although $S$ is an $SU(4)'$ singlet,
we expect $U(1)'$ to be broken below the confining scale through loop
corrections to the scalar potential. In the presence of the condensate
$\langle({\bf 1},{\bf 4}_{-3})({\bf 1},\overline{\bf 4}_{-1})\rangle\simeq \Lambda_{SU(4)}^3$, where $\Lambda_{SU(4)}$ denotes the confining scale of
$SU(4)'$, the scalar potential $V(S)$ of $S$ reads
\begin{equation}\label{eq:V(S)}
 V(S)=\lambda\frac{\Lambda_{SU(4)}^3}{16\pi^2}S+M^2|S|^2+\lambda|S|^4+{\rm h.c.},
\end{equation}
where $\lambda$ is an order one coupling related to $c_{ij}$ and $d_{ij}$ in Eq.~(\ref{eq:SU(4)Lagrangian}), $\lambda'$ is an order one quartic coupling
and the first term is induced by the tadpole diagram shown
in Fig.~\ref{fig:tadpole}.
Minimization of $V(S)$ leads to
$\langle S\rangle\simeq\lambda\Lambda^3_{SU(4)}/(16\pi^2M^2)$. For
$\langle S\rangle\simeq 10^{3}\:{\rm GeV}$ and a typical confining scale
$\Lambda_{SU(4)}\simeq 10^{3}\:{\rm GeV}$, we hence obtain for the mass
$M\simeq 10\:{\rm GeV}$. Therefore, even if $M^2>0$, the $U(1)'$ symmetry
may be spontaneously broken by tadpole terms.

\subsection{$G_{SM}\times SU(3)'\times SU(2)'\times U(1)'$ model}
The chiral and anomaly-free $G_{SM}\times SU(3)'\times SU(2)'\times U(1)'$ model of Ref.~\cite{Foot:1995pa} is simply obtained from the $G_{SM}\times SU(5)'$ model in Sec.~\ref{sec:SU(5)'5H} by restriction to a subgroup. To arrive at this gauge group from $G_{SM}\times SU(5)'$ through SSB, however, one would
require an adjoint Higgs $({\bf 1},{\bf 24})^H$ which does not couple to any
fermions. According to our requirement of having a minimal Higgs content
admitting  only a single Higgs representation in the sterile sector, such a
model would give massless sterile neutrinos with zero mixing between active and sterile neutrinos. A model consistent with our minimality assumption can arise if we start with the $SU(3)'\times SU(2)'\times U(1)'$ subgroup of
$SU(5)'$ as in Ref.~\cite{Foot:1995pa}. Such a model has already been developed, and we have nothing new to add to this case.

\section{$SO(10)\times SO(10)'$ model}\label{sec:SO(10)}
\begin{figure}
\begin{center}
 \includegraphics*[bb=170 309 449 410]{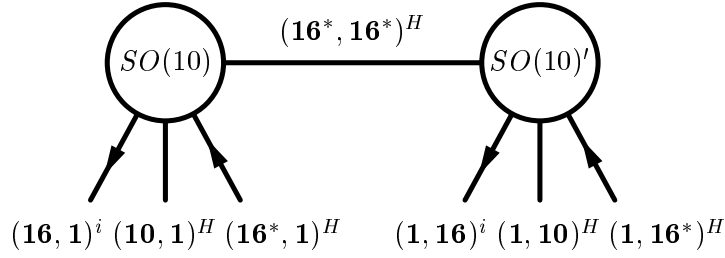}
\vspace*{-1mm}
\caption{\small ``Moose'' or ``quiver'' diagram for the $SO(10)\times SO(10)'$ model. The
effective scalar operator $M_{Pl}^{-1}S_1S_2\sim({\bf 16}^\ast,{\bf 16}^\ast)^H$ links
(with respect to the Yukawa interactions)  as a bispinor the
neighboring gauge groups and thus introduces a nonzero active-sterile neutrino
mixing through the right-handed Majorana sector.
}\label{fig:SO(10)}
\end{center}
\end{figure}

In the models presented above, we have always treated
the right-handed neutrinos $\nu_k^c$, necessary
for the sterile neutrino seesaw mechanism, as
total gauge singlets of the gauge group $G_{SM}\times G'$. We will now
slightly deviate from our general discussion of the $SU(N)$ chains
and consider instead the attractive possibility of unifying all particles, including the right handed neutrinos, into an $SO(10)\times SO(10)'$ product
gauge group. Here, we assume that the SM is embedded into the first group,
{\it i.e.}, $SO(10)\supset G_{SM}$. As the fermionic particle content of this
model we choose the $SO(10)\times SO(10)'$ representations in a symmetrical way as
\begin{equation}\label{eq:SO(10)irreps}
 ({\bf 16},{\bf 1})^i+({\bf 1},{\bf 16})^i,
\end{equation}
where $i=1,2,3$. Like the usual $SO(10)$ models, this gauge theory is chiral and anomaly free. Since all right-handed neutrinos have now been unified into the $SO(10)$ multiplets, we require for the generation of light active and
sterile neutrino masses that the Higgs sector contains two scalars
$S_1\sim({\bf 16}^\ast,{\bf 1})^H$ and $S_2\sim({\bf 1},{\bf 16}^\ast)^H$. These
scalars can generate at the non-renormalizable level Planck-scale suppressed
effective operators
$M_{Pl}^{-1}S_1S_1\sim ({\bf 126}^\ast,{\bf 1})^H$ and
$M_{Pl}^{-1}S_2S_2\sim({\bf 1},{\bf 126}^\ast)^H$ which can supply large Majorana
masses of order $\sim 10^{14}\:{\rm GeV}$ to the right-handed neutrinos.
Similarly, a non-renormalizable operator
$M_{Pl}^{-1}S_1S_2\sim ({\bf 16}^\ast,{\bf 16}^\ast)^H$ is generated, which
transforms as a bispinor under $SO(10)\times SO(10)'$
and appears (with respect to the Yukawa sector) as an effective scalar linking
the two $SO(10)$ gauge groups. The representation content giving rise to
neutrino masses can then be summarized in a
``moose'' \cite{Georgi:1985hf} or ``quiver'' \cite{Douglas:1996sw}
notation in Fig.~\ref{fig:SO(10)}.
Under $SO(10)\supset SU(5)\times U(1)$ and $SO(10)'\supset SU(5)'\times U(1)'$
the representations in Eq.~(\ref{eq:SO(10)irreps}) decompose as\footnote{The
symmetry could be broken along this direction, {\it e.g.}, by 
two extra Higgs fields $({\bf 45},{\bf 1})^H$ and
 $({\bf 1},{\bf 45})^H$.}
\begin{equation}\label{eq:SO(10)decompositions}
 ({\bf 16},{\bf 1})^i=({\bf 10}_1,{\bf 1})^i+
(\overline{\bf 5}_{-3},{\bf 1})^i+({\bf 1}_5,{\bf 1})^i,\quad
({\bf 1},{\bf 16})^i=({\bf 1},{\bf 10}_1)^i+({\bf 1},\overline{\bf 5}_{-3})^i
+({\bf 1},{\bf 1}_5)^i.
\end{equation}
Here, the scalars $S_1$ and $S_2$ decompose according to the conjugate of
the first and the second branching rule, respectively. To generate Dirac
masses in both sectors,
we assume two fundamental Higgs representations
$({\bf 10},{\bf 1})^H$ and $({\bf 1},{\bf 10})^H$ which have under the
decomposition in Eq.~(\ref{eq:SO(10)decompositions}) the branching rules
\begin{equation}
 ({\bf 10},{\bf 1})^H=({\bf 5}_{-2},{\bf 1})^H+(\overline{\bf 5}_2,{\bf 1})^H,\quad
 ({\bf 1},{\bf 10})^H=({\bf 1},{\bf 5}_{-2})^H+({\bf 1},\overline{\bf 5}_2)^H.
\end{equation}
In Eq.~(\ref{eq:SO(10)decompositions}),
the light active and sterile neutrinos are contained in the
$(\overline{\bf 5}_{-3},{\bf 1})^i$ and $({\bf 1},\overline{\bf 5}_{-3})^i$
multiplets, whereas the heavy right-handed neutrinos are identified with the
$SU(5)\times SU(5)'$ singlets
$({\bf 1}_5,{\bf 1})^i$ and $({\bf 1},{\bf 1}_5)^i$.
Up to mass dimension five, the most general Lagrangian relevant for neutrino
masses is found to be
\begin{eqnarray}\label{eq:SO(10)Lagrangian}
 \mathcal{L}_{Y}&=&
a_{ij}({\bf 10},{\bf 1})^H({\bf 16},{\bf 1})^i({\bf 16},{\bf 1})^j
+a'_{ij}({\bf 1},{\bf 10})^H({\bf 1},{\bf 16})^i({\bf 1},{\bf 16})^j
+b_{ij}\frac{S_1^2}{M_{Pl}}({\bf 16},{\bf 1})^i({\bf 16},{\bf 1})^j
\nonumber\\
&+&b'_{ij}\frac{S_2^2}{M_{Pl}}({\bf 1},{\bf 16})^i({\bf 1},{\bf 16})^j
+c_{ij}\frac{S_1S_2}{M_{Pl}}({\bf 16},{\bf 1})^i
({\bf 1},{\bf 16})^j+{\rm h.c.},
\end{eqnarray}
where the coefficients $a_{ij},a_{ij}',b_{ij},b_{ij}'$, and $c_{ij}$ are order
one Yukawa couplings. The two first operators in
Eq.~(\ref{eq:SO(10)Lagrangian}) give rise to Dirac masses of the type
$\sim\nu_i \nu_k^c$ and $\sim\nu_i'{\nu_k'}^c$
for the active and the sterile neutrinos. In the language of the
decompositions in
Eq.~(\ref{eq:SO(10)decompositions}), these Dirac masses arise in
$\mathcal{L}_Y$ from the terms $a_{ij}({\bf 5}_{-2},{\bf 1})^H
(\overline{\bf 5}_{-3},{\bf 1})^i({\bf 1}_{5},{\bf 1})^j$
and $a'_{ij}({\bf 1},{\bf 5}_{-2})^H
({\bf 1},\overline{\bf 5}_{-3})^i({\bf 1},{\bf 1}_{5})^j$, respectively.
When $S_1$ and $S_2$ acquire their VEV's
$\langle S_1\rangle\simeq\langle S_2\rangle\simeq 10^{16}\:{\rm GeV}$,
$S_1$ breaks $SO(10)\rightarrow SU(5)$ and $S_2$ breaks $SO(10)'\rightarrow
SU(5)'$ (the effective bispinor $({\bf 16}^\ast,{\bf 16}^\ast)^H$ acquires its VEV along the $({\bf 1},{\bf 1})$ component under $SU(5)\times SU(5)'$).
Consequently, the third and
fourth terms in Eq.~(\ref{eq:SO(10)Lagrangian}) will generate at the
non-renormalizable level masses for the right-handed neutrinos
of the order $\sim 10^{14}\:{\rm GeV}$. A non-zero mixing between
the active and sterile neutrinos is only introduced in terms of the effective
bispinor $({\bf 16}^\ast,{\bf 16}^\ast)^H$, which couples in the last term
in Eq.~(\ref{eq:SO(10)Lagrangian}) the right-handed neutrinos belonging to
$SO(10)$ with the right-handed neutrinos belonging to $SO(10)'$ by generating
a mixed Majorana mass term of order $\sim 10^{14}\:{\rm GeV}$. After
electroweak symmetry breaking and integrating out the right-handed neutrinos, we thus
arrive at three light active and three light sterile neutrinos with masses in
the (sub)-eV range, which exhibit a nonzero mixing and thus lead to a (3+3) scenario for sterile neutrino oscillations.

Additional Higgs fields, such as $({\bf 45},{\bf 1})^H$ and
$({\bf 1},{\bf 45})^H$ are needed for breaking the $SU(5)\times SU(5)'$
symmetry down to $G_{SM}\times G'$. If the $({\bf 45},{\bf 1})^H$ acquires a VEV along its SM singlet direction $\langle({\bf 45},{\bf 1})^H\rangle\sim
{\rm diag}(a,a,a,b,b)\otimes{\rm i}\sigma^2$, while $\langle({\bf 1},{\bf 45})^H\rangle=0$, we obtain $G_{SM}\times SU(5)'$ of Sec.~\ref{sec:SU(5)'}. If, instead, $\langle({\bf 1},{\bf 45})^H\rangle\sim{\rm diag}(a',a',a',b',b')\otimes{\rm i}\sigma^2$ (along the $SU(3)'\times SU(2)'\times U(1)'$ direction), we have the model of Ref.~\cite{Foot:1995pa}. If, on the other hand,
$\langle({\bf 1},{\bf 45})^H\rangle\sim{\rm diag}(a',a',a',a',b')
\otimes{\rm i}\sigma^2$, we obtain $G_{SM}\times SU(4)'\times U(1)'$ of
Sec.~\ref{sec:SU(4)'}. Thus we see, that all the models described in
Sec.~\ref{sec:SU(5)'} have a natural origin within the $SO(10)\times SO(10)'$
framework.

\section{Other models}\label{sec:othermodels}
\subsection{$G_{SM}\times SU(7)'$ model and its descendants}
\begin{figure}
\begin{center}
 \includegraphics*[bb= 198 136 407 233]{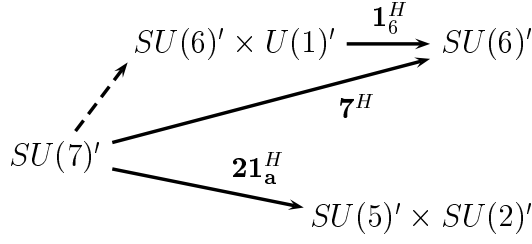}
\vspace*{-2mm}
\caption{\small Symmetry breaking in $SU(7)'$. The solid arrows represent the
symmetry breaking by suitable Higgs representations considered in the text.
The dashed arrow indicates the embedding of $SU(6)'\times U(1)'$ into
$SU(7)'$ as a maximal subgroup.}
\label{fig:SU(7)'}
\end{center}
\end{figure}

In this section, we examine chiral gauge models, where the chain in
Eq.~(\ref{eq:Class(a)}) provides three ``generations'' of antifundamental
representations. The sterile gauge symmetry of these models is 
$SU(7)'$, for which we consider different symmetry breakings as summarized in
Fig.~\ref{fig:SU(7)'}. 

\subsubsection{$SU(7)'$ model}
\label{sec:(1,7)^H}
So far, we have obtained three generations of sterile fermions by taking three
copies of the fermion representations in Eq.~(\ref{eq:Class(a)}) for $N=5$.
As another possibility we shall now consider the case $N=7$, where
the particle content in Eq.~(\ref{eq:Class(a)}) already provides three
antifundamental representations of $SU(7)'$, which we identify with
three sterile fermion generations.
As the gauge group of this model
we therefore have $G_{SM}\times SU(7)'$ with SM singlet fermions transforming
as
\begin{equation}\label{eq:SU(7)'content}
 ({\bf 1},{\bf 21})+({\bf 1},\overline{\bf 7})^i,
\end{equation}
where $i=1,2,3$ is the generation index of the fermions in the
antifundamental representation of $SU(7)'$. Like in the other models discussed
above, we assume in addition to the fermions in
Eq.~(\ref{eq:SU(7)'content}) six right-handed neutrinos
$\nu^c_k$ $(k=1,\ldots ,6)$ which carry zero
$G_{SM}\times SU(7)'$ quantum numbers. To break the $SU(7)'$ symmetry, we suppose a single Higgs representation $S$ which transforms as
$S\sim({\bf 1},{\bf 7})^H$ under $G_{SM}\times SU(7)'$ and acquires its VEV at the TeV scale, {\it i.e.}, $\langle S\rangle\simeq \mathcal{O}({\rm TeV})$. The Lagrangian for neutrino masses of this model reads
\begin{equation}\label{eq:SU(7)'Lagrangian}
  \mathcal{L}_Y=a_{i k}H\ell_i\nu^c_k+b_{ik}S
({\bf 1},\overline{\bf 7})^i\nu^c_k
+c_iS^\ast({\bf 1},{\bf 21})({\bf 1},\overline{\bf 7})^i
+M_{kl}\nu^c_k\nu^c_l+{\rm h.c.},
\end{equation}
where $i,j=1,2,3$ and $k,l=1,\ldots,6$. In Eq.~(\ref{eq:SU(7)'Lagrangian}), the
coefficients $a_{ik}$, $b_{ik}$ and $c_i$ denote complex
$\sim\mathcal{O}(1)$ Yukawa couplings and $M_{kl}\simeq\mathcal{O}(\Lambda)$.
When $S$ acquires its VEV, $SU(7)'$
is broken $SU(7)'\rightarrow SU(6)'$. For $SU(7)'\supset SU(6)'$,
the representations in Eq.~(\ref{eq:SU(7)'content}) and $S$ decompose as
\begin{equation}\label{eq:SU(6)'decompositions}
({\bf 1},{\bf 21})=({\bf 1},{\bf 15})+({\bf 1},{\bf 6}),\quad
({\bf 1},\overline{\bf 7})^i=({\bf 1},\overline{\bf 6})^i+
({\bf 1},{\bf 1})^i.
\end{equation}
where the scalar $({\bf 1},{\bf 7})^H$ decomposes according to the second
equation with ``$i$'' replaced by ``$H$''. After SSB, the interaction
$\sim c_iS^\ast({\bf 1},{\bf 21})({\bf 1},\overline{\bf 7})^i$ in
Eq.~(\ref{eq:SU(7)'Lagrangian}) generates masses
$\sim\mathcal{O}({\rm TeV})$ for $({\bf 1},{\bf 6})$ and
one linear combination of the states
$({\bf 1},\overline{\bf 6})^i$ in Eq.~(\ref{eq:SU(6)'decompositions}) which then decouple. Assuming that also the $G_{SM}\times SU(6)'$ representation
$({\bf 1},{\bf 6})^H$ acquires a mass of order $\sim1\:{\rm TeV}$ and thus decouples from the low energy theory, the leading order beta function
coefficient for $SU(6)'$ is given by $-5{g'}^3/(4\pi^2)$ implying that
$SU(6)'$ is asymptotically free. As a result, all non-trivial fermionic
$SU(6)'$ representations will confine by building condensates of the types
$\langle({\bf 1},\overline{\bf 15})({\bf 1},{\bf 15})\rangle$ and
$\langle({\bf 1},\overline{\bf 6})({\bf 1},{\bf 6})\rangle$ and decouple at
low energies. The $({\bf 1},{\bf 1})^i$ from the $({\bf 1},\overline{\bf 7})^i$, identified with $\nu_i'$, will not acquire any confinement mass, being a singlet of $SU(6)'$. After integrating out
the right-handed neutrinos $\nu^c_k$, the relevant effective Lagrangian for neutrino masses therefore reads like in Eq.~(\ref{eq:effective}), with
$({\bf 1},{\bf 5})^H$ and $({\bf 1},\overline{\bf 5})^i$ respectively
replaced by the representations $({\bf 1},{\bf 21})^H$ and
$({\bf 1},\overline{\bf 7})^i$. As a consequence, we obtain one sterile neutrino per generation and therefore a (3+3) neutrino scheme.

\subsubsection{``Flipped'' $SU(7)'$ model}
We will now consider a model with gauge group
$G_{SM}\times SU(7)'$, which has a fermion sector identical with the model described in Sec.\ref{sec:(1,7)^H}. In contrast to the previous model, however, we assume that the Higgs sector is extended by a Higgs field $S$, which transforms as $S\sim ({\bf 1},{\bf 21})^H$ under $G_{SM}\times SU(7)'$ and acquires a VEV
$\langle S\rangle\simeq\mathcal{O}({\rm TeV})$. We term this model as
``flipped'' $SU(7)'$, in analogy with the flipped $SU(5)'$ model, where an
antisymmetric second rank tensorial Higgs was used. The renormalizable
Lagrangian for neutrino masses now reads
\begin{equation}\label{eq:(1,21)^HLagrangian}
  \mathcal{L}_Y=a_{i k}H\ell_i\nu^c_k+b_kS^\ast
({\bf 1},{\bf 21})\nu^c_k
+c_{ij}S({\bf 1},\overline{\bf 7})^i({\bf 1},\overline{\bf 7})^j
+M_{kl}\nu^c_k\nu^c_l+{\rm h.c.}.
\end{equation}
where $i,j=1,2,3$ and $k,l=1,\ldots,6$. In Eq.~(\ref{eq:(1,21)^HLagrangian}),
the coefficients $a_{ik}$, $b_k$ and $c_{ij}=-c_{ji}$ denote complex order one
Yukawa couplings and $M_{kl}\simeq\mathcal{O}(\Lambda)$.
For a range of parameters, $S$ will acquire a VEV of the skew-symmetric form
$\langle S\rangle\sim{\rm diag}(0,0,0,0,0,1\otimes{\rm i}\sigma^2)$,
thereby breaking $SU(7)'\rightarrow SU(5)'\times SU(2)'$
\cite{li74}. Under $SU(7)'\supset SU(5)'\times SU(2)'$, the representations in Eq.~(\ref{eq:SU(7)'content}) decompose as
\begin{equation}\label{eq:SU(5)'xSU(2)'decompositions}
 ({\bf 1},{\bf 21})=({\bf 1},{\bf 10},{\bf 1})+
({\bf 1},{\bf 5},{\bf 2})+({\bf 1},{\bf 1},{\bf 1}),\quad
({\bf 1},\overline{\bf 7})^i=
({\bf 1},\overline{\bf 5},{\bf 1})^i+({\bf 1},{\bf 1},{\bf 2})^i,
\end{equation}
while $({\bf 1}, {\bf 21})^H$ decomposes according to the first equation. The
interaction
$\sim c_{ij}S({\bf 1},\overline{\bf 7})^i({\bf 1},\overline{\bf 7})^j$ in
Eq.~(\ref{eq:(1,21)^HLagrangian}) generates masses of order
$\sim 1\:{\rm TeV}$ for two linear combinations of the $SU(2)'$ doublets $({\bf 1},{\bf 1},{\bf 2})^i$ in Eq.~(\ref{eq:SU(5)'xSU(2)'decompositions}). Assuming that the $G_{SM}\times SU(5)'\times SU(2)'$
representations $({\bf 1},{\bf 10},
{\bf 1})^H$ and $({\bf 1},{\bf 5},{\bf 2})^H$ have masses of order
$\sim 1\:{\rm TeV}$, the leading order coefficients of the beta
functions for $SU(5)'$ and $SU(2)'$ are respectively given by
$-47{g'_5}^3/(48\pi^2)$ and $-{g'_2}^3/(3\pi^2)$, {\it i.e.}, $SU(5)'$ and
$SU(2)'$ are asymptotically free. As a result, all non-trivial
fermionic representations of $SU(5)'\times SU(2)'$ will confine and decouple
from the low-energy theory, while the trivial $({\bf 1},{\bf 1},{\bf 1})$ representation from $({\bf 1},{\bf 21})$ remains light. After integrating out the
heavy right-handed
neutrinos $\nu^c_k$, the relevant effective Lagrangian for neutrino masses is
then on a form similar to the one given in Eq.~(\ref{eq:effective}), with
$({\bf 1},{\bf 5})^H$ and $({\bf 1},\overline{\bf 5})^i$ respectively
replaced by the representations $({\bf 1},\overline{\bf 21})^H$ and the single
field $({\bf 1},{\bf 21})$ of $G_{SM}\times SU(7)'$. From Eq.~(\ref{eq:SU(5)'xSU(2)'decompositions}) we then conclude that this model gives in total one light sterile neutrino leading to a (3+1) neutrino oscillation scheme.

\subsubsection{$SU(6)'\times U(1)'$ model}
Let us now examine a model which is obtained from the model in
Sec.~\ref{sec:(1,7)^H} by restricting to the (maximal) subgroup
$SU(6)'\times U(1)'$ of $SU(7)'$ (see Fig.~\ref{fig:SU(7)'}).
All sterile fermion representations of this
model thus follow from breaking up the fermion representations
in Sec.~\ref{sec:(1,7)^H} into the representations of
$G_{SM}\times SU(6)'\times U(1)'$. Here, the
representations in Eq.~(\ref{eq:SU(7)'content}) decompose under
$SU(7)'\supset SU(6)'\times U(1)'$ as
\begin{equation}
({\bf 1},{\bf 21})=
({\bf 1},{\bf 15}_2)+({\bf 1},{\bf 6}_{-5}),\quad
({\bf 1},\overline{\bf 7})^i=
({\bf 1},\overline{\bf 6}_{-1})^i
+({\bf 1},{\bf 1}_{-6})^i,
\end{equation}
where $i=1,2,3$. We assume that the SM Higgs sector is extended by a single
Higgs field $S$, which transforms under $G_{SM}\times SU(6)'\times U(1)'$ as
$S\sim(\mathbf{1},\mathbf{1}_6)^H$ and acquires a VEV
$\langle S\rangle\simeq\mathcal{O}({\rm TeV})$. The most general
renormalizable Lagrangian for neutrino masses of this model is
\begin{equation}\label{eq:SU(6)'xU(1)'Lagrangian}
 \mathcal{L}_Y=
 a_{ik}H\ell_i\nu^c_k+b_{ik}S({\bf 1},{\bf 1}_{-6})^i\nu^c_k
+c_{i}S({\bf 1},{\bf 6}_{-5})({\bf 1},\overline{\bf 6}_{-1})^i+M_{kl}\nu_k^c\nu_l^c+{\rm h.c.},
\end{equation}
where $i=1,2,3$, and $k,l=1,\ldots ,6$. In
Eq.~(\ref{eq:SU(6)'xU(1)'Lagrangian}), the quantities $a_{ik}$, $b_{ik}$, and
$c_i$ denote complex order one Yukawa couplings and
$M_{kl}$ the cutoff. When $S$ acquires its VEV at the TeV scale,
the gauge group is
broken $SU(6)'\times U(1)'\rightarrow SU(6)'$. Moreover, $({\bf 1},{\bf 6}_5)$ and one linear combination of the states $({\bf 1},\overline{\bf 6}_{-1})^i$ acquire through SSB a mass $\sim 1\:{\rm TeV}$ and decouple from the low energy theory. Then, the leading order
coefficient of the beta function for $SU(6)'$ is $-5{g'}^3/(4\pi^2)$,
{\it i.e.}, $SU(6)'$ is asymptotically free. Therefore,  all non-singlet $SU(6)'$ representations will decouple by forming the
condensates 
$\langle({\bf 1},\overline{\mathbf{6}}_{-1})^i({\bf 1},\mathbf{6}_{+1})^j
\rangle$, $\langle({\bf 1},\overline{\mathbf{6}}_{5})
({\bf 1},\mathbf{6}_{-5})\rangle$, and
$\langle({\bf 1},\overline{\mathbf{6}}_{-1})^i
({\bf 1},\overline{\mathbf{6}}_{-1})^j({\bf 1},\mathbf{15}_2)\rangle$.
After integrating out the right-handed neutrino singlets $\nu^c_k$, the effective Lagrangian for neutrino masses becomes similar to the Lagrangian in
Eq.~(\ref{eq:effective}) with $({\bf 1},{\bf 5})^H$ and
$({\bf 1},\overline{\bf 5})^i$ respectively replaced by $({\bf 1},{\bf 1}_6)^H$
and $({\bf 1},{\bf 1}_{-6})^i$.  In this model, we therefore obtain
one light sterile neutrino per generation leading to a (3+3) neutrino oscillation scheme.

\subsection{Models in the $G_{SM}\times SU(3)'$ chain}
\subsubsection{$SU(3)'$ model}
\label{sec:SU(3)'}
The smallest gauge group which allows a chiral and anomaly-free gauge theory
in the class shown in Eq.~(\ref{eq:Class(b)})
is $SU(3)'$, which 
has seven fields in the antifundamental representation and one in the
symmetric second rank tensor representation. As total gauge group let us now take
$G_{SM}\times SU(3)'$ with the extra fermion representations transforming as
\begin{equation}\label{eq:SU(3)'content}
({\bf 1},{\bf 6})+({\bf 1},\overline{\bf 3})^i,
\end{equation}
where $i=1,\ldots ,7$. In order to break $SU(3)'$ we assume a Higgs field $S$
which transforms as $S\sim ({\bf 1},{\bf 3})^H$ under $G_{SM}\times SU(3)'$
and acquires a VEV $\langle S\rangle\simeq\mathcal{O}({\rm TeV})$, which can
always be written as $\langle S\rangle=(0,0,|s|)^T$. We
furthermore add six right-handed neutrinos $\nu_k^c$ $(k=1,\ldots ,6)$ which are total singlets of $G_{SM}\times SU(3)'$. The most general renormalizable Lagrangian
for neutrino masses is then given by
\begin{equation}\label{eq:SU(3)'Lagrangian}
\mathcal{L}_Y=
a_{\alpha k}H\ell_\alpha\nu^c_k+b_{ik}S({\bf 1},\overline {\bf 3})^i\nu^c_k
+c_iS^\ast({\bf 1},{\bf 6})({\bf 1},\overline{\bf 3})^i+
M_{kl}\nu^c_k\nu^c_l+{\rm h.c.},
\end{equation}
where $\alpha=1,2,3$ and $i=1,\ldots ,7$. In Eq.~(\ref{eq:SU(3)'Lagrangian}), the coefficients $a_{\alpha k}$, $b_{ik}$, and $c_i$ are $\sim \mathcal{O}(1)$ Yukawa couplings and $M_{kl}$ is of order the cutoff scale.
When $S$ acquires its VEV a the TeV scale, the gauge group is broken as $SU(3)'\rightarrow SU(2)'$. With this embedding, the decomposition of the fermion
representations in Eq.~(\ref{eq:SU(3)'content}) for $SU(3)'\supset SU(2)'$ read
\begin{equation}\label{eq:SU(2)'decompositions}
 ({\bf 1},{\bf 6})=
 ({\bf 1},{\bf 3})+({\bf 1},{\bf 2})+({\bf 1},{\bf 1}),\quad
 ({\bf 1},\overline{\bf 3})^i=({\bf 1},{\bf 2})^i
+({\bf 1},{\bf 1})^i,
\end{equation}
where $i=1,\ldots ,7$ and $S$ decomposes here as
$({\bf 1},{\bf 3})^H=({\bf 1},{\bf 2})^H+({\bf 1},{\bf 1})^H$. In Eq.~(\ref{eq:SU(2)'decompositions}), the
representations $({\bf 1},{\bf 2})$, $({\bf 1},{\bf 1})$, one linear
combination of the states $({\bf 1},{\bf 2})^i$ and one linear combination
of the singlets $({\bf 1},{\bf 1})^i$ will acquire masses of order
TeV through the interaction
$c_iS({\bf 1},{\bf 6})({\bf 1},\overline{\bf 5})^i$ in Eq.~(\ref{eq:SU(3)'Lagrangian}). Assuming that the $G_{SM}\times SU(2)'$ representation
$({\bf 1},{\bf 2})^H$ decouples by obtaining a mass
$\sim\mathcal{O}({\rm TeV})$, the leading order coefficient of the beta
function for $SU(2)'$ becomes $-{g_2'}^3/(4\pi^2)$, {\it i.e.}, $SU(2)'$ is
asymptotically free.
Consequently, in Eq.~(\ref{eq:SU(2)'decompositions}),
the two massless linear combinations of the states
$({\bf 1},{\bf 2})^i$ will decouple by forming condensates of the types
$\langle ({\bf 1},{\bf 2})({\bf 1},{\bf 2})\rangle$.
After integrating out the right-handed neutrinos
$\nu_k^c$, the effective Lagrangian generating neutrino
masses can therefore be written as in Eq.~(\ref{eq:effective}), with
$({\bf 1},{\bf 5})^H$ replaced by $({\bf 1},{\bf 3})^H$ and
$({\bf 1},\overline{\bf 5})^i$ replaced by $({\bf 1},\overline{\bf 3})^i$.
In total, this model therefore leads to six light sterile neutrinos identified
in Eq.~(\ref{eq:SU(2)'decompositions})
with six linear combinations of the singlets $({\bf 1},{\bf 1})^i$
and we hence we obtain a (3+6) model for sterile neutrino oscillations.

\subsubsection{$SU(2)'\times U(1)'$ model}
We consider now the anomaly-free and chiral gauge theory which is obtained
from the previous model in Sec.~\ref{sec:SU(3)'} by restricting to the
subgroup $G_{SM}\times SU(2)'\times U(1)'$. The fermion content of this
model then results from breaking up the fermion representations in
Sec.~\ref{sec:SU(3)'} into the representations 
of $G_{SM}\times SU(2)'\times U(1)'$. In particular, the multiplets in Eq.~(\ref{eq:SU(3)'content}) decompose under $SU(3)'\supset SU(2)'\times U(1)'$ as
\begin{equation}\label{eq:SU(2)'xU(1)'decompositions}
 ({\bf 1},{\bf 6})=
 ({\bf 1},{\bf 3}_2)+({\bf 1},{\bf 2}_{-1})+({\bf 1},{\bf 1}_{-4}),\quad
({\bf 1},\overline{\bf 3})^i=({\bf 1},{\bf 2}_{-1})^i+({\bf 1},{\bf 1}_{+2})^i,
\end{equation}
where $i=1,\ldots ,7$ and the subscript denotes the $U(1)'$ charge. Moreover,
we assume seven right-handed neutrinos $\nu_k^c$ $(k=1,\ldots ,7)$ which are total singlets under $G_{SM}\times SU(2)'\times U(1)'$. We suppose
that $U(1)'$ is spontaneously broken by a single Higgs fields $S$ which
transforms under $G_{SM}\times SU(2)'\times U(1)'$ as
$S\sim ({\bf 1},{\bf 1}_{-2})$ and acquires a VEV
$\langle S\rangle\simeq \mathcal{O}({\rm TeV})$. Here, the Lagrangian relevant
for neutrino masses reads
\begin{eqnarray}\label{eq:SU(2)'xU(1)'Lagrangian}
 \mathcal{L}_Y&=&
 a_{\alpha k}H\ell_\alpha \nu_k^c+b_{ik}S({\bf 1},{\bf 1}_{+2})^i\nu_k^c
+c_iS^\ast({\bf 1},{\bf 1}_{-4})({\bf 1},{\bf 1}_{+2})^i\nonumber\\
&+&d_{ij}S^\ast({\bf 1},{\bf 2}_{-1})^i({\bf 1},{\bf 2}_{-1})^j
+d'_iS^\ast({\bf 1},{\bf 2}_{-1})({\bf 1},{\bf 2}_{-1})^i
+M_{kl}\nu^c_k\nu^c_l+{\rm h.c.},
\end{eqnarray}
where $\alpha=1,2,3$ and $i,j,k,l=1,\ldots,7$.
In Eq.~(\ref{eq:SU(2)'xU(1)'Lagrangian}), the
coefficients $a_{\alpha k}$, $b_{ik}$, $b'_i$, $c_{ij}$, and $d_i$ denote
$\sim\mathcal{O}(1)$ Yukawa couplings and $M_{kl}\simeq\mathcal{O}(\Lambda)$.
When the scalar $S$ acquires its VEV at the TeV scale,
the gauge group is broken as $SU(2)'\times U(1)'\rightarrow SU(2)'$. The Yukawa interactions with $S$ will therefore generate masses of order
$\sim 1\:{\rm TeV}$ for all eight $SU(2)'$ doublets,
$({\bf 1},{\bf 1}_{-4})$, and one linear combination of the states
$({\bf 1},{\bf 1}_2)^i$ in Eq.~(\ref{eq:SU(2)'xU(1)'decompositions}),
which hence decouple from the low energy theory. The leading order coefficient of the beta
function for $SU(2)'$ then becomes $-{g'}^3/(4\pi^2)$ and $SU(2)'$ is
asymptotically free. After integrating out the heavy right-handed neutrinos
$\nu^c_k$, we therefore obtain two light sterile neutrinos per generation,
leading to a (3+6) scheme for sterile neutrino oscillations.

\section{Summary and conclusions}\label{sec:summary}
In this paper, we have presented a family of chiral gauge models which would
protect the masses of sterile neutrinos. A ``sterile'' gauge symmetry enables us to realize a seesaw mechanism for the SM singlet neutrinos. Our main
motivation for studying this class of models is to provide an explanation for the LSND neutrino oscillation result, in conjunction with the solar and atmospheric neutrino data. The naturally light sterile neutrinos of our models
can find application as candidates for warm dark matter, which could also provide an understanding of the observed anomalously large radio pulsar velocities exceeding
$\sim 500\:{\rm km/s}$ (``pulsar kicks'') \cite{ful}. There is yet another possible application. With the recent revision in the value of the Boron production rate $S_{17}(O)$, the total solar neutrino flux observed by the SNO experiment seems to indicate a deficit of active
neutrinos by $\sim 12\%$.
This deficit may be understood via a small admixture of light
sterile neutrinos \cite{deHolanda:2003tx} which would be easily provided by
our models.

The class of models we have constructed has a natural embedding
in an $SO(10)\times SO(10)'$ unified theory as discussed in
Sec.~\ref{sec:SO(10)}. The $SU(5)'$ model of Sec.~\ref{sec:SU(5)'5H} as well as its descendants of Secs.~\ref{sec:SU(5)'10H}, \ref{sec:Sp(4)'}, and
\ref{sec:SU(4)'} can all be neatly embedded into $SO(10)\times SO(10)'$. We have also constructed models based on an $SU(7)'$ gauge symmetry, which provides
{\it three} light sterile neutrinos naturally, and models based on an $SU(3)'$ gauge symmetry with fermions in the sextet representation.

The most direct test of our models will be a confirmation of the LSND oscillation data by the ongoing MiniBooNE experiment. Reactor neutrino disappearance experiments, as well as neutrinoless double beta decay experiments should be sensitive to the existence of sterile neutrinos \cite{Babu:2003is}. It should be noted that the standard big bang nucleosynthesis \cite{wal} will be affected by
the presence of $\nu'$, however, there are ways around it, such as by assuming primordial
lepton asymmetry \cite{foot} or with low reheating temperature
\cite{Gelmini:2004ah}. The bound on neutrino masses from recent cosmological
data \cite{el} may also be alleviated by such a lepton asymmetry
\cite{Hannestad:2003ye}. Although the scale of new physics is of order TeV in our models, testing them at colliders will be challenging, since the extended gauge sector has no direct couplings to the SM sector. One possible signature
is the invisible decay of
the SM Higgs boson $H$, as its mixing with the Higgs field $S$ used for sterile gauge symmetry breaking can be substantial. Invisible decays such as
$H\longrightarrow W'W'$ can then occur, with $W'$, the gauge bosons of the sterile gauge symmetry, decaying into sterile fermions.

\section*{Acknowledgments}
We would like to thank Ts.~Enkhbat for useful comments and discussions.
This work is supported in part by DOE Grant \# DE-FG02-04ER46140 and an award
from the Research Corporation.

\end{document}